\documentclass[10pt, conference, compsocconf]{IEEEtran}

\usepackage[utf8]{inputenc}

\usepackage{graphicx}
\usepackage{todonotes}
\usepackage{paralist}
\usepackage{booktabs}
\usepackage{flushend}

\newcommand{\squeeze}{}

\title{Which requirements artifact quality defects are automatically detectable? A case study}
\begin{document}

\author{
\IEEEauthorblockN{Henning Femmer}
\IEEEauthorblockA{Institut für Informatik\\
Technische Universit\"at M\"unchen, Germany\\
femmer@in.tum.de
}
\and\IEEEauthorblockN{Michael Unterkalmsteiner, Tony Gorschek}
\IEEEauthorblockA{
Software Engineering Research Lab, \\
Blekinge Institute of Technology, Sweden\\
\{mun,tgo\}@bth.se
}
}

%
\maketitle
\IEEEpeerreviewmaketitle              

\begin{abstract}
\textbf{[Context:]}
The quality of requirements engineering artifacts, e.g. requirements 
specifications, is acknowledged to be an important success factor for projects. Therefore, many companies spend 
significant amounts of money to control the quality of their RE artifacts. 
To reduce spending and improve the RE artifact quality, methods were proposed 
that combine manual quality control, i.e. reviews, with automated approaches.
\textbf{[Problem:]} So far, we have seen various approaches to automatically detect certain aspects in RE artifacts. However, we still lack an overview what can and cannot be automatically detected.
\textbf{[Approach:]} Starting from an industry guideline for RE artifacts, we classify 166 existing rules for RE artifacts along various categories to discuss the share and the characteristics of those rules that can be automated. For those rules, that cannot be automated, we discuss the main reasons.
\textbf{[Contribution:]} We estimate that 53\% of the 166 rules can be checked automatically either perfectly or with a good heuristic. Most rules need only simple 
techniques for checking. The main reason why some rules resist automation is due to 
imprecise definition.
\textbf{[Impact:]} By giving first estimates and analyses of automatically 
detectable and not automatically detectable rule violations, we aim to provide 
an overview of the potential of automated methods in requirements quality 
control.
\end{abstract}


\IEEEpeerreviewmaketitle

\section{Introduction}

Requirements Engineering (RE) artifacts play a central role in many systems and 
software engineering projects. Due to that central role, the quality of RE 
artifacts is widely considered a success factor, both in academia, e.g. by 
Boehm~\cite{Boehm1988} or Lawrence~\cite{Lawrence2001}, and also by 
practitioners~\cite{MendezFernandez2015}.

As a result, companies invest heavily into quality 
control of RE artifacts. Since RE artifacts are written mostly in natural 
language~\cite{Luisa2004}, quality control is usually applied manually, e.g.\ in the form of manual reviews. 
However, besides all of its advantages, manual quality control is slow, 
expensive and inconsistent, heavily dependent on the competence of the 
reviewer. One obvious approach to address this is combining manual reviews 
with automated approaches. The goal of a so-called \emph{phased 
inspection}~\cite{Knight1993a,FemmerQAProcess16} is to reduce the effort in 
manual reviews and to improve the review results by starting into the review 
with a better (e.g. readable) artifact.

Therefore, various authors have focused on automatically detecting 
quality defects, such as ambiguous language 
(i.a.~\cite{Wilson1997,Fabbrini2001,lucassen2016improving,Femmer2016}) or 
cloning~\cite{Juergens2010}. 
However, it is still an open question to what degree quality defects can be 
detected automatically or require human expertise (i.e. manual work). In 
previous work~\cite{Femmer2016}, we took a bottom-up perspective by qualitatively analyzing which of the quality review 
results could be automatically detected. 

\noindent\textbf{Research Goal:} In this work, we take a top-down perspective 
by focusing on requirements writing guidelines from a large company. Furthermore, we 
systematically classify and quantify which proportion of the rules can be 
automated.

\section{Related Work}
Researchers and practitioners have been working on supporting quality assurance 
with automated methods (at least) since the end of the 
1990's~\cite{Wilson1997}. We want to give only a brief, non-exhaustive summary 
here. Please refer to our previous work~\cite{Femmer2016} for a more detailed 
analysis.

\noindent\textbf{Defect types:} Most works in this area focus on the detection of various forms of ambiguity, e.g.~\cite{Fabbrini2001,Fantechi2003,Knauss2009b,Genova2011}. Other works try to detect violations of syntactic~\cite{Juergens2010} or even semantic duplications~\cite{Falessi2013}. Other works focus on correct classifications~\cite{winkler2016} or on the question whether an instance follows given structural guidelines, e.g. for user stories~\cite{lucassen2016improving} or for use cases~\cite{Alchimowicz2011}. 

\noindent\textbf{Criteria:} The aforementioned works used different sets of 
criteria. Most prominently are definitions of ambiguity~\cite{Berry2003}, 
previously summarized lists of criteria~\cite{berry2006new}, or requirements 
standards~\cite{Femmer2016,femmer14}.

\noindent\textbf{Techniques:} So far, various techniques have been applied, including machine learning~\cite{winkler2016,Yang2011} and ontologies~\cite{Korner2009a}. However, Arendse and Lucassen~\cite{arendse2016toward} hypothesize that we might not need sophisticated methods for most aspects of quality. In this paper, we provide data regarding this hypothesis.
All in all, few works have tried to take a different viewpoint and understand 
what \emph{cannot} be automatically checked. In previous 
work~\cite{Femmer2016}, we approached this question in a qualitative manner, 
by looking not at definitions, but at instances of defects. We did not quantify 
the portion of automatically discoverable defects, since this depends heavily 
on the requirements at hand (which defects does an author introduce and 
a reviewer find?).

\textbf{Research Gap:} Various authors have shown how to automatically detect 
individual quality defects. In previous work~\cite{Femmer2016} 
we qualitatively analyzed which requirements quality defects can be detected.
In this work, we provide first evidence, based on requirements 
writing rules used in a large organization, on the proportion between 
automatically\,/\,not automatically detectable requirements quality 
issues.

\section{Study Design}
We conducted this study in a research collaboration with the Swedish Transport 
Administration (STA), the government agency responsible for planning, 
implementing and maintaining long-term rail, road, shipping and aviation 
infrastructure in Sweden. In particular, we studied their requirements 
guidelines that were developed by editors who review and quality assure 
specifications. A total of 129 rules were analyzed in this paper. While our 
long-term goal in this research collaboration, is described in more detail 
elsewhere~\cite{unterkalmsteiner_requirements_2017}, 
the specific research goal of this paper 
is to \emph{characterize requirements writing rules with respect to their 
potential to be automatically checked from the viewpoint of a 
requirements quality researcher in the context of an industrial requirements quality control process}.

\noindent From this goal definition we derive our research questions:
\begin{enumerate}
\item[RQ1:] How many rules for natural language requirements specifications can be automated?
\item[RQ2:] To what degree can rules be categorized into groups and to what 
degree can these groups be eligible for automation?
\item[RQ3:] What information is required to automatically detect rule 
violations?
\item[RQ4:] Which rules resist automation and why?
\end{enumerate}

\subsection{Rule classification}\label{sec:ruleclass}
\begin{table*}[t]
	\caption{Classification schema with rule examples}
	\begin{center}
		\begin{tabular}{lp{6cm}lllp{2.5cm}p{1.54cm}}
			\toprule
			ID & Rule & Type & Context & Scope & Necessary information & 
			Detection \mbox{accuracy}\\
			\midrule
			160 & The term ``function'' shall be used instead of the term 
			``functionality''. & Lexical & Anywhere & Word/Phrase & 
			Lemma / Dictionary & Deterministic\\
			56 & Requirements shall start with the subject. & Grammatical & 
			Requirement & Sentence & Parse tree & Heuristic (h)\\
			78 & Text consisting of a definition shall be preceded with the 
			identifier ``Definition:''. & Structural & Requirement & Section & 
			Lemma / Dictionary & Heuristic (m)\\
			81 & If a functional requirement is supplemented with additional 
			information to clarify how the requirement can be met, the 
			additional information must be formulated as a separate 
			requirement. & 
			Semantic & Requirement & Section & Domain model & Heuristic (l)\\
			24 & References to other documents in the specification are done by 
			reference to the document title. & Structural & Anywhere & Global & 
			Regular expressions, Document list & Deterministic\\
			50 & Requirements must be understandable independently, i.e. the 
			subject must be indicated in the respective requirements (the 
			subject must not be only defined in the section title). & Semantic 
			& Requirement & Sentence & POS tags & Heuristic (h)\\  
			54 & The introductory section of the specification shall not 
			contain any requirements. & - & - & - & - & Not detectable\\
			\bottomrule
		\end{tabular}
	\end{center}
	\label{tab:classificationschema}
\end{table*}
A lack of classification schema for requirements writing rules prompted us to 
formulate the following schema (see Tbl.~\ref{tab:classificationschema}).

\subsubsection{Rule type}
We distinguish between the lexical, grammatical, structural and semantic rule 
type (see rules 160, 56, 78 and 81 in Tbl.~\ref{tab:classificationschema}). A 
lexical rule refers to constraints on the use of certain terms or 
expressions that may induce ambiguity, reduce understandability or readability. 
Similarly, a grammatical rule refers to constraints on sentence composition. A 
structural rule refers to the form in which information is presented and 
formatted. Finally, a semantic rule refers to constraints on the text content 
and meaning. 

\subsubsection{Rule context}
We introduced this dimension to characterize in which context of the 
requirements specification the rule is relevant. An appropriate automated check 
flags only violations that occur in the correct context, e.g. in 
requirements (if they are separated from informative text), figures, tables, 
references, headings, enumerations, comments.

\subsubsection{Information scope}
This dimension describes the scope that needs to be considered in order to 
decide whether the rule is violated or not. We defined five levels: 
word/phrase, sentence, section, document and global. For example, to check rule 
56 in Tbl.~\ref{tab:classificationschema}, it is enough to inspect a sentence. 
However, rule 24 requires access to information that is not in the requirements 
specification, hence we classified it as global information scope. This 
characterization provides indication that can be used to estimate the relative 
required effort to implement the automated check of the rule.

\subsubsection{Necessary information}
This dimension describes NLP-based and domain-specific information needed to 
detect rule violations. NLP-based information refers to language and document 
structure, such as Part-of-Speech (POS) tags, lemmas and word stems, 
morphological tags, parse trees and meta-data on formatting. Domain-specific 
information is only available in the specific domain in which the rules apply, 
e.g. lists of referenced documents or a domain model / ontology. For example, 
rule 50 in Tbl.~\ref{tab:classificationschema} can be decided with POS tags 
while rule 56 requires a parse tree that indicates where the subject is 
positioned in the sentence. 

\subsubsection{Detection accuracy}
\begin{figure}
	\begin{center}
		\includegraphics[width=.7\columnwidth]{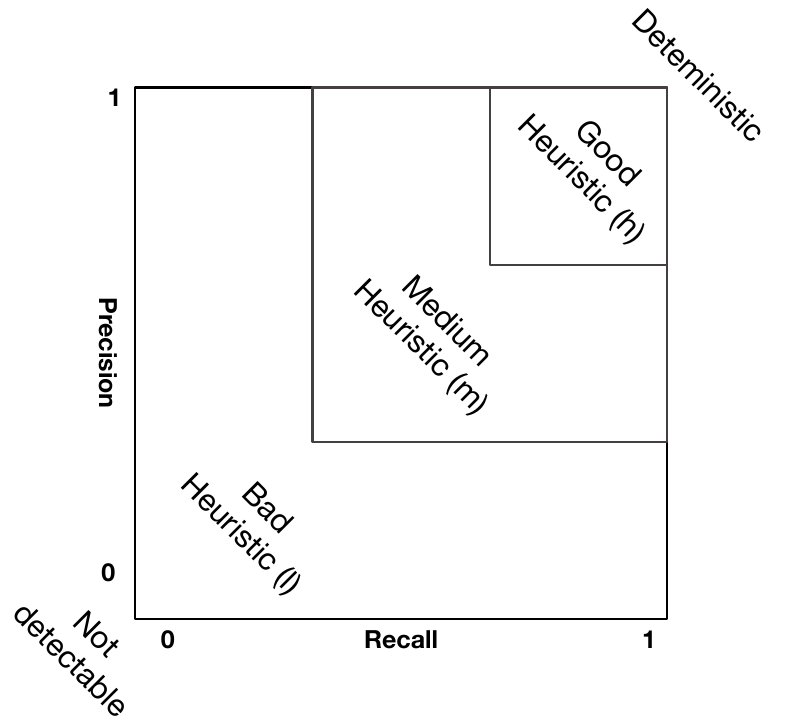}
		\caption{The categories of detection accuracy as used in this 
		study}
		\label{fig:accuracy}
	\end{center}
\end{figure}

This dimension provides a rough estimate, based on the experiences of 
previous work~\cite{femmer14}, on the expected accuracy for 
detecting rule violations. We have defined a five-level scale, illustrated in 
Fig.~\ref{fig:accuracy}, spanning from deterministic, i.e. 100\% detectable, to not detectable at all. Good heuristics feature both high 
recall and precision, while bad heuristics always trade-off between precision 
and recall. For example, while assigning POS tags is a probabilistic algorithm, 
we classified rule 50 in Tbl.~\ref{tab:classificationschema} as a good 
heuristic since this particular problem has been solved before, with 
demonstrably high precision and recall. We classified rule 81, on the other hand, as bad heuristic since, while conceptually feasible, we lack an 
accurate solution, i.e. a technique to extract a domain model and use that to 
determine whether a requirement statement contains supplemental information. 
Then, there are also rules that we do not expect to be automatically detectable 
at all (rule 54), because they turn out to be challenging, even in manual reviews. We classified these not automatically detectable rules along main reasons (categories resulted from previous work~\cite{Femmer2016}, see Tbl.~\ref{tab:reasons}).

\subsection{Data Collection, Classification and Analysis}
We received a total of 192 writing rules from STA, of which we filtered 
unapproved rule ideas (63), resulting in 129 original rules. In case a rule 
contained discernible sub-rules, we split them up to facilitate 
the classification, resulting in 166 classified rules.
We then developed an initial version of the classification schema illustrated 
in Section~\ref{sec:ruleclass}. While all
dimensions and the categories for type and detection accuracy were defined 
a-priori, the categories for context, scope and necessary information 
were identified during the classification process. During this first workshop 
we classified 39 rules, stabilizing the schema and fostering our shared 
understanding. Then, the second author proceeded to classify 
the remaining 127 rules alone.  The first author sampled 20 rules from this 
set, independently classified them and calculated the inter-rater agreement 
($\kappa=0.79$) which is considered substantial~\cite{landis_measurement_1977}. 
The first author then reviewed all 127 rules, marked those where he disagreed, 
and finally consolidated all classifications with the second author in a second 
workshop. 

We then used the classifications of accuracy for RQ1, the type, context and scope for RQ2, the necessary information for RQ3, and the reasons for RQ4.

\section{Results}

\subsection*{RQ1: How many rules for natural language requirements specifications can be automated?}
\begin{figure}[htbp]
\begin{center}
\includegraphics[width=.9\columnwidth]{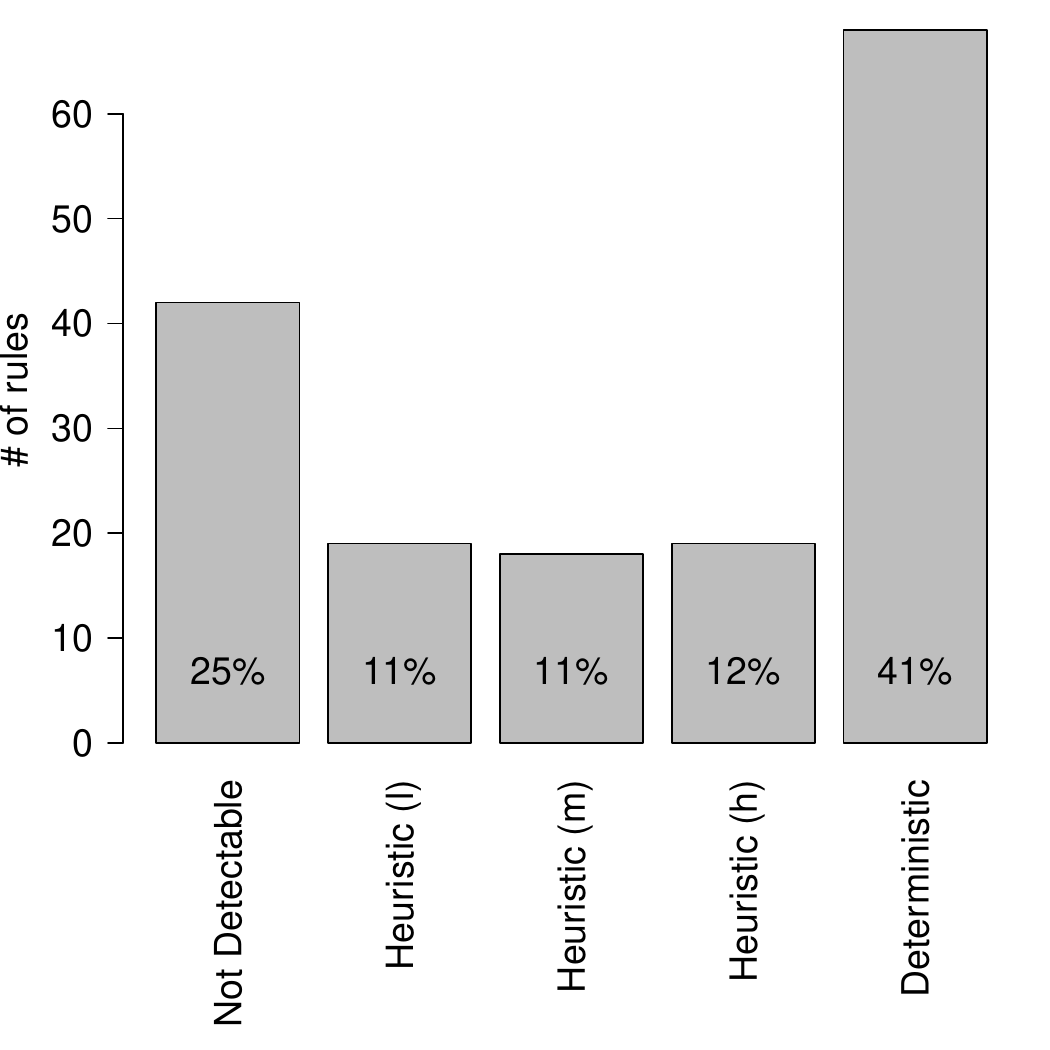}
\caption{Frequency of rules falling into one of the detection accuracy 
categories.}\squeeze
\label{fig:rq1}
\end{center}
\end{figure}

In Fig.~\ref{fig:rq1}, we show the results from classifying the estimated 
detection accuracy of the rules. 
We estimate that 41\% of the rules can be deterministically checked, 
meaning that an algorithm finds each violation. 34\% of the 
rules are heuristic, with 12\% of high accuracy, and 11\% of medium and low accuracy. We 
estimate that the remaining 25\% cannot be checked at the current state of art 
and at the current state of the rule definitions.

\noindent\emph{Discussion:}
Whether rules can be automatically detected is not a binary question. In fact, 
it depends on the context. However, most rules we can put into a certain 
category, indicating their potential to be automatically checked. We were 
surprised by the large number of rules that can be automated. This indicates 
the potential for automation, as we will discuss in future work.

\begin{figure}[htbp]
\begin{center}
\includegraphics[width=\columnwidth]{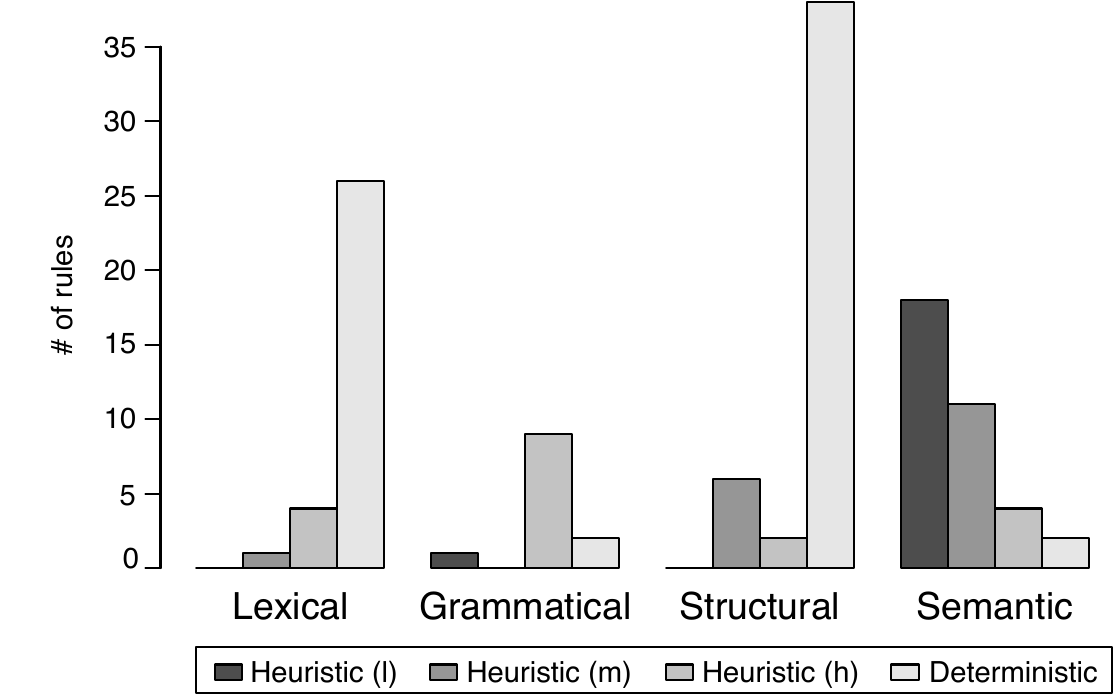}
\caption{Estimated detection accuracy for each category.}\squeeze
\label{fig:rq2}
\end{center}
\end{figure}

\begin{figure}[htbp]
\begin{center}
\includegraphics[width=\columnwidth]{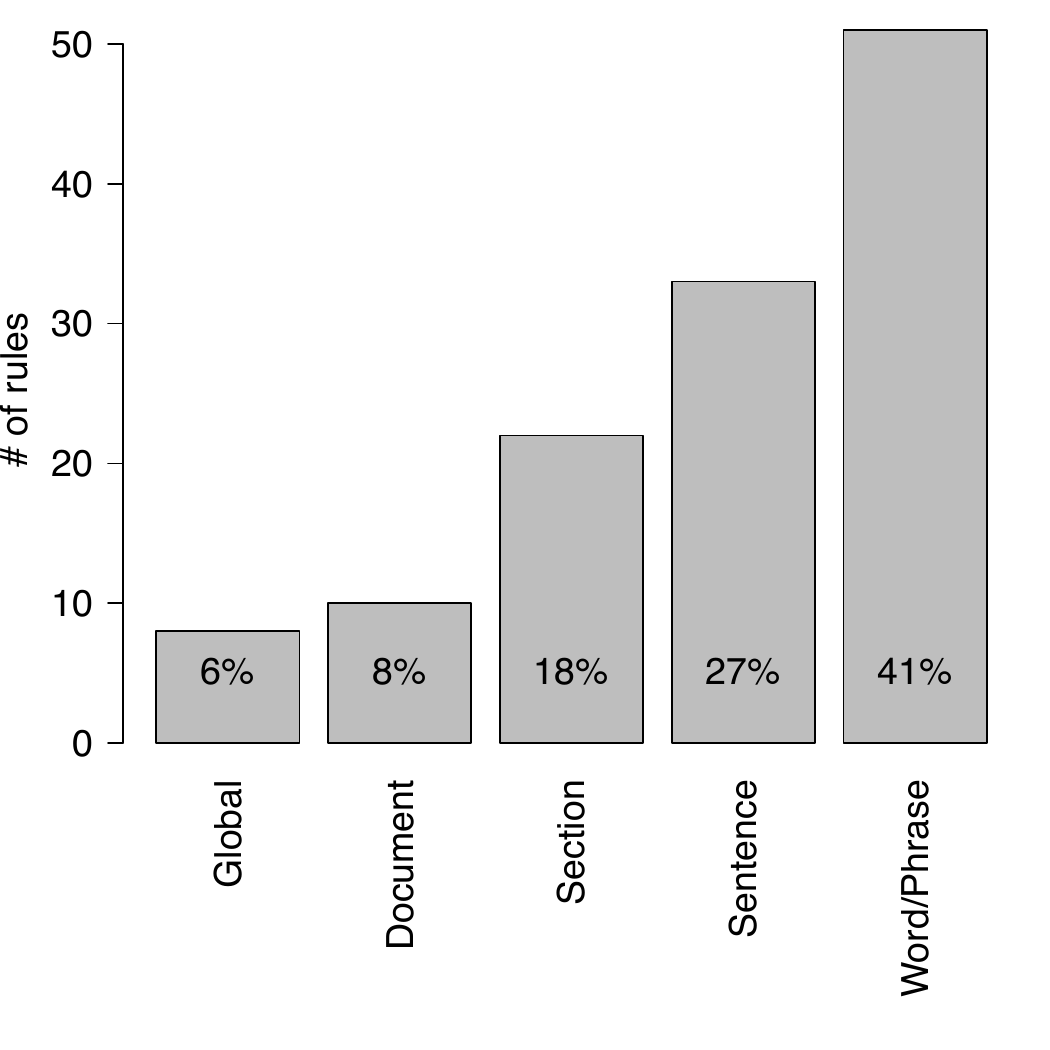}\vspace{-1em}
\caption{Distribution of the scope of the automatically detectable rules.}\vspace{-1em}
\label{fig:rq21}
\end{center}
\end{figure}

\begin{figure}[htbp]
\begin{center}
\includegraphics[width=\columnwidth]{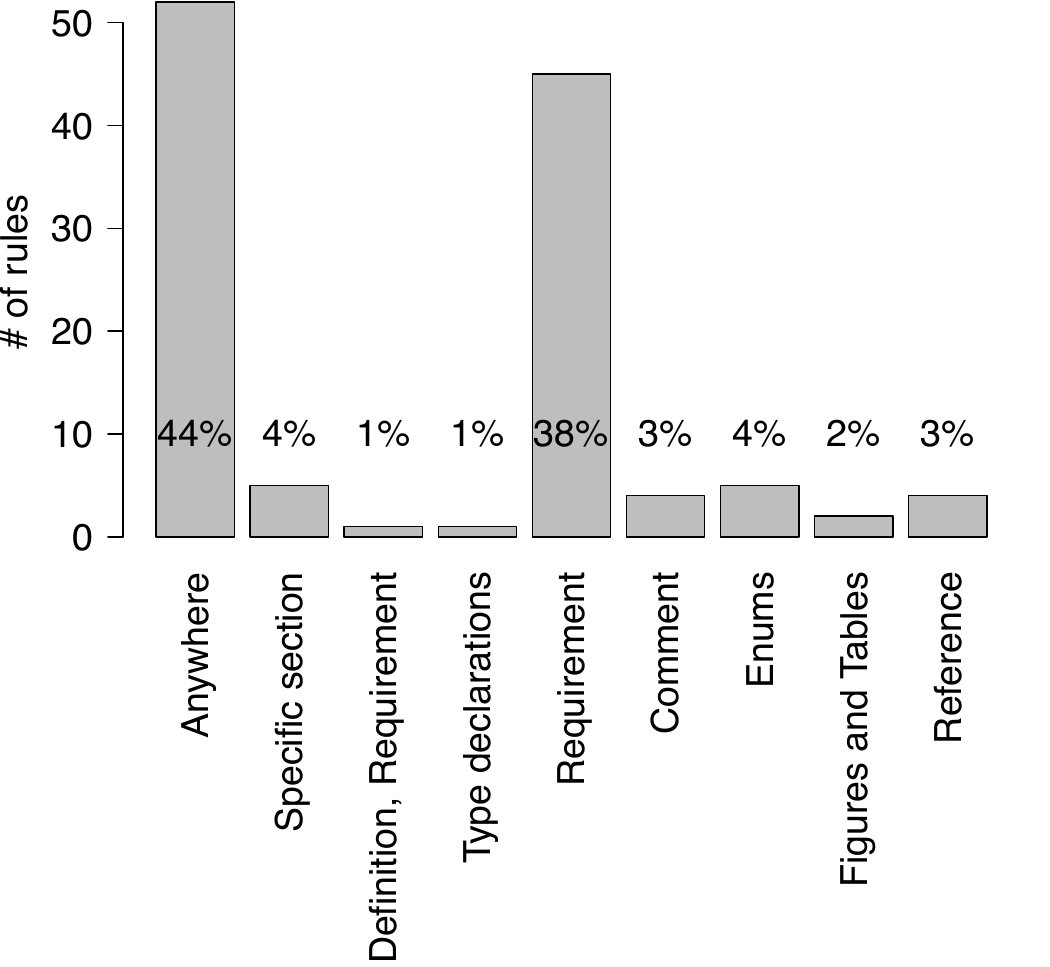}
\caption{Context of the automatically detectable rules.}\squeeze
\label{fig:rq22}
\end{center}
\end{figure}

\subsection*{RQ2: To what degree can rules be categorized into groups and to 
what degree can these groups be eligible for automation?}
In Fig.~\ref{fig:rq2}, we show the results from classifying the automatically 
detectable rules by their type and estimated detection accuracy. The results 
indicate an estimated high detection accuracy for structural and lexical rules, 
medium accuracy for grammatical rules, and medium to low accuracy for semantic 
rules. 
Fig.~\ref{fig:rq21} shows that most rules are at the level of words or phrasing or at the level of sentences. Lastly, Fig.~\ref{fig:rq22} shows that most rules hold anywhere or specifically concern the requirements of the RE artifact.

\noindent\emph{Discussion:}
The further a rule goes into semantic aspects, the harder it is 
to detect violations. For structural rules, e.g.\ where a certain piece of  
information should be placed, there are a few rules for which violations are 
difficult to check automatically. For example, to understand whether a certain 
text should be tagged as a requirement requires context understanding. 
We describe further reasons for rules not being automatically detectable in RQ4.

\subsection*{RQ3: What information is required to automatically detect rule 
violations?}
To understand what techniques are required to automatically detect violations 
of guideline rules, we classified each rule with the required information for 
this rule. Each required information then leads to a certain technique. For 
example, if the lemmas of the words are required, we obviously need a 
lemmatization technique. 
Tbl.~\ref{tab:techniques} shows the results for this analysis. The three most 
common techniques are the following: In 47\% of the cases, lemmatization is 
required to detect a violation of a rule. In a further 35\% of cases only the 
pure text and regular expressions are needed. Next, formatting information is 
required in 22\% of the cases.

\noindent\emph{Discussion:} This analysis supports the hypothesis of Arendse and Lucassen~\cite{arendse2016toward} that in most cases, we do not need sophisticated methods to detect violations of rules.

\begin{table}[htbp]
\caption{Frequency of Required Information (Multiple Selections)}
\begin{center}
\begin{tabular}{l c c}
\toprule
Information & Occurrences & Share of Rules \\
\midrule
Lemmas / Dictionaries & 58 & 47 \%\\
Pure Text (Reg. Expression) & 43 & 35 \%\\
Formatting & 27 & 22 \%\\
Domain Models & 11 & 9 \%\\
Part of Speech Tags & 11 & 9 \%\\
Lists of [X] & 8 & 6 \%\\
Morphology & 5 & 4 \%\\
Parse Trees & 3 & 2 \%\\
Word Stems& 3 & 2 \%\\
Tokens / Sentences & 3 & 2 \%\\
Named Entities & 1 & 1 \%\\
\bottomrule
\end{tabular}\squeeze
\end{center}
\label{tab:techniques}
\end{table}%

\subsection*{RQ4: Which rules resist automation and why?}
When analyzing the \emph{not automatically detectable} rules of RQ1, the 
reasons were distributed as shown in Tbl.~\ref{tab:reasons} (classification 
extends previous work~\cite{Femmer2016}).
The major reason was, in our studied case, that the rules themselves are still 
imprecise or unclear. Examples for this are rules such as \emph{"Requirements 
must be accurate, unambiguous, 
comprehensive, consistent, modifiable, traceable."} (this was one 
single rule) or \emph{"Requirements should contain enough information."} These 
rules cannot be checked either manually or automatically. One could even
argue that they convey little value. Such imprecise or unclear rules are the reason for 81\% of the not 
automatically detectable rules (see Tbl.~\ref{tab:reasons}). In 12\% of the 
cases, an automation would need profound domain knowledge to automatically 
detect a violation. An example is 
that requirements about certain system parts must first state that 
these parts exist. However, to understand which parts this refers to, 
we would need to know the domain. This means that only domain experts 
can manually detect violations to these rules. In one case, 
respectively, the rule requires deep semantic understanding of the text (e.g. to detect logical contradictions written in natural language in different paragraphs), the 
system or even the process scope.

\begin{table}[htbp]
\caption{Share of Reasons that Prevent Automated Detection}
\begin{center}
\begin{tabular}{l c c}
\toprule
Reason & Frequency & Share \\
\midrule
$R_1$: Rule unclear or imprecise  & 34 & 81 \% \\
$R_2$: Deep semantic text understanding & 1 & 2 \% \\
$R_3$: Profound domain knowledge & 5 & 12 \% \\
$R_4$: System scope knowledge & 1 & 2 \% \\
$R_5$: Process status knowledge & 1 & 2 \% \\\midrule
Sum & 42 & 100 \% \\
\bottomrule
\end{tabular}\squeeze
\end{center}
\label{tab:reasons}
\end{table}%

\noindent\emph{Discussion:} 
Deep computational problems do not seem to be the major cause for why we see no 
chance in checking a certain rule, rather imprecise rules themselves. 

\section{Discussion}

\subsection{Share of automatically detectable defects}
In our study, we found that a substantial number of requirements writing rules 
can be automatically checked. This is a top-down perspective and as such helps 
to quantify the share of defects that can be automatically detected. However, 
this does not necessarily transfer to the share of defects found in reviews. 
This is for the following reasons: First, defects created by requirements 
engineers are not equally distributed over the guideline rules. Furthermore, 
the defects introduced by requirements engineers very much depend on the 
individual person, company, and project. Second, defects discovered by 
reviewers are not necessarily equally distributed over the guideline rules. 
Therefore, we argue to consider both perspectives, i.e. the share of defects 
based on guidelines and the share of defects existing in practice, when 
discussing the potential of automated requirements quality assurance.

\subsection{The 100\%-Recall Argument}
There is an ongoing debate in the scientific community whether automated
checks in quality assurance need 100\% recall to be useful in 
practice. Some authors (i.a.~\cite{Berry2012,Kiyavitskaya2008,Tjong2013}) argue 
that if an approach does not achieve perfect recall, this leads to either the 
reviewer does not check the rule anymore, which would lead to unchecked defects,
or the reviewer has to go through the whole document anyways, and thus, the 
automated analysis has no benefits.
We disagree with this view for two reasons. First, we argue that in industrial 
practice, reviewers rarely go through the artifact rule by rule. Therefore, 
there is no such thing as \emph{omitting a certain rule}. Reviewers
see the guidelines rather as a supporting instrument, and thus 
anything that reminds them of certain rules, increases the quality. 
Our second argument also refers to the status quo today. The best automated 
quality support that is widely used are spell and grammar checks. Both 
do not have 100\% recall. So, if recall is a problem, why do we use spell 
and grammar checks every day?
In our experience from introducing automated analyses at various 
companies in industry, practitioners were more worried about precision 
than recall. They are convinced of the value (\emph{"Anything 
helps!"}), and care more for acceptance with the end users. Here, the core 
aspect is usability in the form of few false positives, ergo: precision (cf.\ 
also similar discussing in static code analysis~\cite{ayewah2008}). 

\subsection{Threats to Validity}
There are two major threats to validity. Regarding internal validity, we 
classified the rules according detection accuracy. We did so because it was not 
feasible within the scope of this work to do a precision \& recall analysis for 
each guideline rule. However, the first author has been translating guideline 
rules into automated analyses for 4 years. Thus, we are confident that the 
results reflect the real precision and recall after implementation. In 
addition, we created rough categories to gain an overview, not a precise 
analysis for each rule. To evaluate this aspect, we independently classified a 
subset of 10\% of the rules and calculated a weighted Cohen's kappa of the 
resulting classification ($\kappa = 0.79$). This agreement fosters our 
confidence in the resulting classification.

The second threat relates to external validity. Since we analyzed a 
large guideline used at STA, we do not know whether the results generalize from 
this partner. We have, however, previously informally checked a guideline from 
another industry partner in a different domain. Here we came to the 
same share of not automatically detectable rules (25\%). Future work 
should broaden the study to different guidelines.

\section{Research Agenda}
The current paper provides an estimation of the extent to which industrial 
requirements quality rules can be automatically checked. We plan to continue 
our research as follows.

\noindent\textbf{Complete the rule classification.}
34 of the studied rules were imprecise or unclear. Unfortunately, the authors 
of the writing guidelines were not available for feedback during the course of 
this study. We want to deepen our understanding on the nature of the 
imprecision of these rules. In addition, we 
had no access regarding the relevance, value, and frequency of violations of the rules. 
This could provide insights how rules that can be automatically 
checked potentially contribute to review effort reduction.
In addition, the classification scheme used in RQ2 was beneficial for this 
study and worked fine regarding the first three categories (lexical, 
grammatical, structural). However, the scheme created some 
discussion around the semantic category. The reason is that most rules 
intertwine semantic and syntactic aspects: Since requirements artifacts are not 
automatically compiled like code, the point of syntactic rules is only to 
prevent semantic issues. Therefore, future work should extend this 
classification scheme to clarify this aspect, e.g.\ by decoupling the two 
aspects.

\noindent\textbf{Implement and statically validate rules.}
We have already begun to implement some rules that are based on dictionary 
lookups using an existing requirements smell detection 
framework~\cite{Femmer2016}. 
While most of the rules can be implemented with simple techniques, we also plan 
to experiment with more advanced NLP techniques where we expect challenges in 
the detection accuracy. For example, violations to rule 81 in 
Table~\ref{tab:classificationschema} could be detected by using topic models 
enhanced with domain knowledge~\cite{andrzejewski2009incorporating}: 
requirements that contain distant topics or several closely related topics 
indicate candidates for rule violations.
To validate the implemented rules, we can exploit the fact that at 
STA, the rules were developed based on experience, i.e.\ there exist versions of 
requirements that contain rule violations. We can fine-tune and validate the 
detection against this set. We also plan to provide an 
analysis of the potential benefits of using automated requirements quality 
control. 
To achieve this, we analyze historic requirements (where the 
current rules were not applied) and study the effort spent on discussing and 
repairing these violations.

\noindent\textbf{Validation in Use.}
We plan to evaluate the efficiency and effectiveness of automated 
requirements quality assurance in use, i.e. in the environment of STA with the 
support of their requirements editors. One important question to answer is 
whether we can control the number of false positives, a crucial aspect for the 
adoption of tool support in industry that has also been observed in other 
areas, such as static bug detection~\cite{ayewah2008}. 

\noindent\textbf{Repository for requirements writing rules.}
Finally, we, as a community, should establish a repository of precise general and  
validated requirements rules. 
Such a repository can be created by replicating the work proposed in this paper 
in different contexts and, at the same time, advance the techniques for detecting rule 
violations.

\section{Conclusions}
It is unclear what proportion of quality defects can be automatically 
detected. Therefore, in this work, we classify rules from a large, fine-grained 
requirements writing guideline from one of our industry partners. 
The results indicate that a surprisingly large proportion of rules (41\%) can 
be automatically analyzed. 53\% can be analyzed deterministically or with a good heuristic. One reason for this was that these rules contain 
many structural rules, which require just an analysis of formatting information 
or pure text. If we take also those rules into account where we have a 
medium heuristic, we could even tackle 64\% of the rules. 
However, our analysis also shows that 36\% of the rules have no or little 
chance to be automated.
While being just first evidence, this analysis indicates that there is a 
substantial proportion of guideline rules (our intermediate for quality 
defects) 
that can be automatically checked. However, the analysis also indicates that 
there is little hope that we can completely replace manual reviewing with 
automated reviews. Combining automated and manual quality assurance, as 
proposed by others~\cite{Knight1993a}, and also 
ourselves~\cite{FemmerQAProcess16} could be the promising compromise.

\section*{Acknowledgements}
This work was performed within the project Q-Effekt and ERSAK; it was funded by 
the German Federal Ministry of Education and Research (BMBF) under grant no. 
01IS15003 A-B and by the Swedish Transport Administration. The authors assume 
responsibility for the content. The authors thank Jonas Eckhardt for comments on an earlier draft of this paper.

\bibliographystyle{IEEEtran}
\bibliography{references}

\begin{thebibliography}{10}
\providecommand{\url}[1]{#1}
\csname url@samestyle\endcsname
\providecommand{\newblock}{\relax}
\providecommand{\bibinfo}[2]{#2}
\providecommand{\BIBentrySTDinterwordspacing}{\spaceskip=0pt\relax}
\providecommand{\BIBentryALTinterwordstretchfactor}{4}
\providecommand{\BIBentryALTinterwordspacing}{\spaceskip=\fontdimen2\font plus
\BIBentryALTinterwordstretchfactor\fontdimen3\font minus
  \fontdimen4\font\relax}
\providecommand{\BIBforeignlanguage}[2]{{%
\expandafter\ifx\csname l@#1\endcsname\relax
\typeout{** WARNING: IEEEtran.bst: No hyphenation pattern has been}%
\typeout{** loaded for the language `#1'. Using the pattern for}%
\typeout{** the default language instead.}%
\else
\language=\csname l@#1\endcsname
\fi
#2}}
\providecommand{\BIBdecl}{\relax}
\BIBdecl

\bibitem{Boehm1988}
B.~W. Boehm and P.~N. Papaccio, ``Understanding and controlling software
  costs,'' \emph{IEEE Transactions on Software Engineering}, vol.~14, no.~10,
  pp. 1462--1477, 1988.

\bibitem{Lawrence2001}
B.~Lawrence, K.~Wiegers, and C.~Ebert, ``The top risks of requirements
  engineering,'' \emph{{IEEE Software}}, pp. 62--63, 2001.

\bibitem{MendezFernandez2015}
D.~{M{\'e}ndez Fern{\'a}ndez} and S.~Wagner, ``Naming the pain in requirements
  engineering: A design for a global family of surveys and first results from
  germany,'' \emph{Information and Software Technology}, vol.~57, pp. 616--643,
  2015.

\bibitem{Luisa2004}
L.~Mich, F.~Mariangela, and P.~L. Novi~Inverardi, ``Market research for
  requirements analysis using linguistic tools,'' \emph{Requirements
  Engineering Journal}, vol.~9, no.~1, pp. 40--56, 2004.

\bibitem{Knight1993a}
J.~C. Knight and E.~A. Myers, ``{An improved inspection technique},''
  \emph{Communications of the ACM}, vol.~36, no.~11, pp. 51--61, 1993.

\bibitem{FemmerQAProcess16}
H.~Femmer, B.~Hauptmann, S.~Eder, and D.~Moser, ``Quality assurance of
  requirements artifacts in practice: A case study and a process proposal,'' in
  \emph{PROFES}, 2016, pp. 506--516.

\bibitem{Wilson1997}
W.~M. Wilson, L.~H. Rosenberg, and L.~E. Hyatt, ``{Automated analysis of
  requirement specifications},'' in \emph{ICSE}, 1997, pp. 161--171.

\bibitem{Fabbrini2001}
F.~Fabbrini, M.~Fusani, S.~Gnesi, and G.~Lami, ``{An automatic quality
  evaluation for natural language requirements},'' in \emph{REFSQ}, 2001.

\bibitem{lucassen2016improving}
G.~Lucassen, F.~Dalpiaz, J.~M.~E. van~der Werf, and S.~Brinkkemper, ``Improving
  agile requirements: the quality user story framework and tool,''
  \emph{Requirements Engineering}, vol.~21, no.~3, pp. 383--403, 2016.

\bibitem{Femmer2016}
H.~Femmer, D.~{M{\'e}ndez Fern{\'a}ndez}, S.~Wagner, and S.~Eder, ``Rapid
  quality assurance with requirements smells,'' \emph{Journal of Systems and
  Software}, vol. 123, pp. 190--213, 2017.

\bibitem{Juergens2010}
E.~Juergens, F.~Deissenboeck, M.~Feilkas, B.~Hummel, B.~Schaetz, S.~Wagner,
  C.~Domann, and J.~Streit, ``{Can Clone Detection Support Quality Assessments
  of Requirements Specifications?}'' in \emph{ICSE}, 2010.

\bibitem{Fantechi2003}
A.~Fantechi, S.~Gnesi, G.~Lami, and A.~Maccari, ``{Application of linguistic
  techniques for Use Case analysis},'' \emph{Requirements Engineering}, vol.~8,
  no.~3, pp. 161--170, 2002.

\bibitem{Knauss2009b}
E.~Knauss, D.~L{\"{u}}bke, and S.~Meyer, ``{Feedback-Driven Requirements
  Engineering : The Heuristic Requirements Assistant},'' in \emph{ICSE}, 2009.

\bibitem{Genova2011}
G.~G{\'{e}}nova, J.~M. Fuentes, J.~Llorens, O.~Hurtado, and V.~Moreno, ``{A
  framework to measure and improve the quality of textual requirements},''
  \emph{Requirements Engineering}, vol.~18, no.~1, pp. 25--41, sep 2011.

\bibitem{Falessi2013}
D.~Falessi, G.~Cantone, and G.~Canfora, ``Empirical principles and an
  industrial case study in retrieving equivalent requirements via natural
  language processing techniques,'' \emph{IEEE Transactions on Software
  Engineering}, vol.~39, no.~1, pp. 18--44, 2013.

\bibitem{winkler2016}
J.~Winkler and A.~Vogelsang, ``Automatic classification of requirements based
  on convolutional neural networks,'' in \emph{3rd International Workshop on
  Artificial Intelligence for Requirements Engineering (AIRE)}, 2016.

\bibitem{Alchimowicz2011}
B.~Alchimowicz, J.~Jurkiewicz, J.~Nawrocki, and M.~Ochodek, ``{Towards
  use-cases benchmark},'' \emph{Software Engineering Techniques}, 2011.

\bibitem{Berry2003}
D.~M. Berry, E.~Kamsties, and M.~M. Krieger, ``{From Contract Drafting to
  Software Specification : Linguistic Sources of Ambiguity},'' 2003.

\bibitem{berry2006new}
D.~M. Berry, A.~Bucchiarone, S.~Gnesi, G.~Lami, and G.~Trentanni, ``A new
  quality model for natural language requirements specifications,'' in
  \emph{REFSQ}, 2006, pp. 1--12.

\bibitem{femmer14}
H.~Femmer, D.~{M{\'e}ndez Fern{\'a}ndez}, E.~Juergens, M.~Klose, I.~Zimmer, and
  J.~Zimmer, ``Rapid requirements checks with requirements smells: Two case
  studies,'' in \emph{International Workshop on Rapid Continuous Software
  Engineering}, 2014, pp. 10--19.

\bibitem{Yang2011}
H.~Yang, A.~D. Roeck, V.~Gervasi, A.~Willis, and B.~Nuseibeh, ``{Analysing
  anaphoric ambiguity in natural language requirements},'' \emph{Requirements
  Engineering}, vol.~16, no.~3, pp. 163--189, 2011.

\bibitem{Korner2009a}
S.~J. K{\"{o}}rner and T.~Brumm, ``{Natural Language Specification Improvement
  With Ontologies},'' \emph{International Journal of Semantic Computing},
  vol.~3, no.~4, pp. 445--470, 2009.

\bibitem{arendse2016toward}
B.~Arendse and G.~Lucassen, ``Toward tool mashups: Comparing and combining {NLP
  RE} tools,'' in \emph{3rd International Workshop on Artificial Intelligence
  for Requirements Engineering (AIRE)}, 2016, pp. 26--31.

\bibitem{unterkalmsteiner_requirements_2017}
M.~Unterkalmsteiner and T.~Gorschek, ``Requirements quality assurance in
  industry: why, what and how?'' in \emph{{REFSQ}}, 2017.

\bibitem{landis_measurement_1977}
J.~R. Landis and G.~G. Koch, ``The {Measurement} of {Observer} {Agreement} for
  {Categorical} {Data},'' \emph{Biometrics}, vol.~33, no.~1, pp. 159--174,
  1977.

\bibitem{Berry2012}
D.~M. Berry, R.~Gacitua, P.~Sawyer, and S.~F. Tjong, ``{The case for dumb
  requirements engineering tools},'' \emph{Lecture Notes in Computer Science},
  vol. 7195 LNCS, pp. 211--217, 2012.

\bibitem{Kiyavitskaya2008}
N.~Kiyavitskaya, N.~Zeni, L.~Mich, and D.~M. Berry, ``{Requirements for tools
  for ambiguity identification and measurement in natural language requirements
  specifications},'' \emph{Requirements Engineering}, vol.~13, no.~3, pp.
  207--239, jul 2008.

\bibitem{Tjong2013}
S.~F. Tjong and D.~M. Berry, ``{The design of SREE - A prototype potential
  ambiguity finder for requirements specifications and lessons learned},'' in
  \emph{REFSQ}, 2013, pp. 80--95.

\bibitem{ayewah2008}
N.~Ayewah, D.~Hovemeyer, J.~D. Morgenthaler, J.~Penix, and W.~Pugh, ``Using
  static analysis to find bugs,'' \emph{IEEE Software}, vol.~25, no.~5, pp.
  22--29, 2008.

\bibitem{andrzejewski2009incorporating}
D.~Andrzejewski, X.~Zhu, and M.~Craven, ``Incorporating domain knowledge into
  topic modeling via dirichlet forest priors,'' in \emph{{ICML}}.\hskip 1em
  plus 0.5em minus 0.4em\relax ACM, 2009, pp. 25--32.

\end{thebibliography}

\end{document}